\documentclass[sigconf,letter, screen]{acmart}
 
\usepackage{booktabs} \usepackage[utf8]{inputenc}
\usepackage{enumerate}
\usepackage{ifmtarg}
\usepackage{soul}
\usepackage{tabularx,ragged2e}
\usepackage[export]{adjustbox}
\usepackage{mdframed}
\usepackage[autostyle=true]{csquotes}
\usepackage{xspace}

\usepackage{colortbl}
\usepackage{array, booktabs, boldline} \usepackage{cellspace}
\usepackage{sidecap}

\usepackage{listings}

\usepackage{color}
\definecolor{gray}{rgb}{0.4,0.4,0.4}
\definecolor{darkblue}{rgb}{0.0,0.0,0.6}
\definecolor{cyan}{rgb}{0.0,0.6,0.6}

\lstdefinelanguage{XML}
{
  morestring=[b]",
  morestring=[s]{>}{<},
  morecomment=[s]{<?}{?>},
  stringstyle=\color{black},
  identifierstyle=\color{darkblue},
  keywordstyle=\color{cyan},
  morekeywords={xmlns,version,type}}

\setcopyright{rightsretained}

\newif\ifcomments
\newif\ifchanges
\commentsfalse\changesfalse
\commentstrue\changestrue

\ifcomments
\newcommand{\commentbox}[1]{\noindent\framebox{\parbox{0.98\linewidth}{#1}}}

\newcommand{\acomment}[2]{\ \\ \fbox{\parbox{0.98\linewidth}{{\sc #1}: #2}}}
\newcommand{\mcomment}[2]{{\color{blue}(#1)}\footnote{#1: #2}}                                 \else
\newcommand{\commentbox}[1]{}
\newcommand{\mcomment}[2]{}
\newcommand{\acomment}[2]{}
\fi

\ifchanges

\setul{}{0.2mm}
\setstcolor{red}

\else

\fi

 \newcommand{\tzm}[1]{\mcomment{TZ}{#1}}
 \newcommand{\ggm}[1]{\mcomment{GG}{#1}}
 \newcommand{\fvm}[1]{\mcomment{FV}{#1}}
 \newcommand{\alm}[1]{\mcomment{AL}{#1}}
 \newcommand{\spm}[1]{\mcomment{SP}{#1}}
 \newcommand{\jsm}[1]{\mcomment{JS}{#1}}

 \newcommand{\illtis}{\textsc{Iltis}\ }
 \newcommand{\Illtis}{\illtis}
 
 \renewcommand{\implies}{\rightarrow}
 
\newcommand{\anon}{$\blacksquare\blacksquare\blacksquare\blacksquare\blacksquare$}
\newcommand{\anonymize}[2]{#1}

\renewcommand{\tzm}[1]{}
\renewcommand{\ggm}[1]{}
\renewcommand{\fvm}[1]{}
\renewcommand{\alm}[1]{}
\renewcommand{\spm}[1]{}
\renewcommand{\jsm}[1]{}

\setcopyright{acmlicensed}
\acmPrice{15.00}
\acmDOI{10.1145/3197091.3197095}
\acmYear{2018}
\copyrightyear{2018}
\acmISBN{978-1-4503-5707-4/18/07}
\acmConference[ITiCSE'18]{23rd Annual ACM Conference on Innovation and Technology in Computer Science Education}{July 2--4, 2018}{Larnaca, Cyprus}

\begin{document}
\title[Introduction to \Illtis\hspace{-0.8mm}: An Interactive, Web-Based System for Teaching Logic]{Introduction to \Illtis\hspace{-1mm}: \\ An Interactive, Web-Based System for Teaching Logic}

\anonymize{
  \titlenote{The authors acknowledge the financial support from the state of North Rhine-Westphalia
  in the form of funding for the improvement of university education as well as the financial
  support by a Fellowship for Innovation in Digital University Education by the state of North
  Rhine-Westphalia and the Stifterverband held by the last author.}
  \author{Gaetano Geck}
  \affiliation{    \institution{TU Dortmund}
    \country{Germany}
          }
  \email{gaetano.geck@tu-dortmund.de}

  \author{Artur Ljulin}
  \affiliation{    \institution{TU Dortmund}
   \country{Germany}
          }
  \email{artur.ljulin@tu-dortmund.de}

  \author{Sebastian Peter}
  \affiliation{    \institution{TU Dortmund}
    \country{Germany}
          }
  \email{sebastian.peter@tu-dortmund.de}

  \author{Jonas Schmidt}
  \affiliation{    \institution{TU Dortmund}
    \country{Germany}
          }
  \email{jonas2.schmidt@tu-dortmund.de}

  \author{Fabian Vehlken}
  \affiliation{    \institution{TU Dortmund}
    \country{Germany}
          }
  \email{fabian.vehlken@tu-dortmund.de}

  \author{Thomas Zeume}
  \affiliation{    \institution{TU Dortmund}
    \country{Germany}
          }
  \email{thomas.zeume@tu-dortmund.de}

    \renewcommand{\shortauthors}{G. Geck et al.}
}{}

\begin{abstract}
Logic is a foundation for many modern areas of computer science. In artificial intelligence, as a basis of database query languages, as well as in formal software and hardware verification  --- modelling scenarios using logical formalisms and inferring new knowledge are important skills for going-to-be computer scientists.

The \Illtis project aims at providing a web-based, interactive system that supports teaching logical methods. In particular the system shall (a) support to learn to model knowledge and to infer new knowledge using propositional logic, modal logic and first-order logic, and (b) provide immediate feedback and support to students. 

This article presents a prototypical system that currently supports the above tasks for propositional logic. First impressions on its use in a second year logic course for computer science students are reported.
\end{abstract}

\ccsdesc[500]{Applied computing~Interactive learning environments}
\ccsdesc[500]{Theory of computation~Logic}

\keywords{Logic, interactive learning environment}

\maketitle

\section{Introduction}\label{section:introduction}
Logical formalisms play an important role in many modern computer science applications. In artificial intelligence, knowledge is often modelled by logical formulas and logical inference mechanisms are used for inferring new (though implicit) knowledge. The foundation of many modern database query languages such as SQL, Cypher and XQuery are logical formalisms. In software and hardware verification, desirable properties are often specified by logic-based specification languages such as the temporal logics LTL and CTL, and the correctness of systems with respect to such specifications is verified using logical procedures.

Learning logical formalisms and in particular logical modelling is therefore inevitable for computer science students. Typical lectures introducing  logic to computer science students focus on teaching how to model computer science scenarios by logical means  and how to infer new knowledge from such a representation.

For example, the first weeks of the introductory logic lecture for computer scientists at \anonymize{the University of Dortmund}{\anon} cover
\begin{enumerate}[(a)]
  \item how to model real world scenarios by propositional formulas,
  \item the transformation of formulas into an adequate normal form, and
  \item the inference of new knowledge (represented as propositional formulas) using an inference mechanism\ggm{I hesitate to call it \emph{inference mechanism}, although, in a sense, this is right.} such as propositional resolution.
\end{enumerate}
Afterwards a similar process is introduced for modal logic and first-order logic, thereby equipping students with the means to later learn similar      logical languages from application areas such as artificial intelligence, verification, or databases by themselves.

Each of the tasks (a)--(c) is simple and can be performed quite mechanically. Yet, as most advanced logical topics require fluency in these tasks, it is essential that students practice each of them and see how they play together in solving problems.

After learning the steps (a)--(c) for propositional logic, students are expected to be able to solve problems such as the following.
\begin{example}\label{example:exercise_running_example}
After carefully investigating a faulty software system, Julia has found the following dependencies between the three components of the system:

\begin{enumerate}[(1)]
\item If the database is faulty, then so is the back end.
\item The back end is only faulty if both the database and the user interface are faulty.
\item Not all three components are faulty.
\end{enumerate}

Julia concludes that the database is correct. Can you verify her conclusion by modelling the situation in propositional logic and inferring Julia's conclusion using propositional resolution? \qed
\end{example}

\paragraph{Goal of this Work and Contribution} The goal of the \Illtis project is to develop a web-based, interactive system that (1) supports teaching the modelling process (a)--(c) for propositional logic, modal logic and first-order logic, and (2) provides immediate, didactically valuable feedback and assistance when required. The system shall allow for easy inclusion of further logics, other typical tasks, and additional feedback mechanisms.

In this article we present a web-based prototypical application and its underlying framework that support the
modelling process for propositional logic outlined above. Each step required to solve the problem
from Example \ref{example:exercise_running_example} is implemented as a task. The framework\ggm{Not \emph{so} urgent, but also here (and in following places) it could, if desired, be made more precise what is offered by the framework and what implemented by the application} allows to specify exercises by combining such tasks in a flexible way in XML. Tasks can provide feedback using a generic mechanism, which is implemented in detail to provide feedback for a task where students have to model statements by propositional
formulas. A preliminary evaluation of the framework has been performed in the winter term 2017/2018.

We acknowledge that students usually struggle rather with modal and first-order logic than with propositional logic. However, focusing on propositional logic so far has allowed us to create a robust architecture and gather experience in how to design tasks and useful feedback mechanisms. In the conclusion we sketch our vision for the future of the project.

Technically the \Illtis framework is implemented in Java using the Google Web Toolkit (GWT), offering a webpage user interface based solely on HTML and Javascript. We intend to publish the source code as open source as soon as a stable version is available.

\paragraph{Related work}
We concentrate on web-based systems as they are accessible to most students.
Several existing systems cover some of the tasks we are aiming at. The \emph{LogEX} system allows for training the transformation of  propositional formulas \cite{LodderHJ15, LodderH11}. The inference of new knowledge using calculi that are close to natural inference is supported by many systems, see e.g. \cite{Sieg07, HuertasHLM11, EhleHL15, GasquetSS11a}. Resolution, which is used in the introduction to logic in \anonymize{Dortmund}{\anon}, is to the best of our knowledge only supported by the \emph{AELL} system \cite{HuertasHLM11} (which is not publicly available and only has support for the Spanish and Catalan language).
In \emph{Tarski's World} students can learn how to evaluate first-order logic in a 3D-world. A playful but prototypical approach towards topics in an introductory logic course is taken in \cite{SchaferHLSBZ13}.

Digital tools are also used in some interactive logic books. For example, the teaching environment \cite{Fricke12} allows for transforming textual statements into logical formulas by a mark-and-replace technique. Many small interactive tasks can be found in the interactive book \cite{Vellemann10} as well as on the support websites of \emph{Power of Logic}~\cite{WassermanHH12}. An inspiration for presenting models for modal formulas (i.e. Kripke structures) can be found in~\cite{Kirsling15}.

An overview over further, also older systems, can be found in~\cite{Huertas11}.

In summary, some of the aspects \Illtis aims at are covered by other systems. Yet, an integration of these systems seems to be impracticable or even impossible due to technological diversity.

\paragraph{Organization}
The system is introduced from a teacher's perspective in Section \ref{section:user_perspective}. After presenting the architecture and components of the \Illtis system in Section \ref{section:architecture}, we report on the feedback generation for writing propositional formulas in Section \ref{section:feedback}. In Section \ref{section:evaluation} we discuss first class room experiences. We conclude in Section \ref{section:conclusion}.
 
\section{The \Illtis System: The Teacher's Perspective}\label{section:user_perspective}

\begin{figure*}

\begin{minipage}{.485\linewidth}
  \begin{mdframed}
\scriptsize
\lstset{language=XML}
\makeatletter
\lst@AddToHook{OnEmptyLine}{\vspace{-0.88\baselineskip}}
\makeatother
\begin{lstlisting}[mathescape=true]
<?xml version="1.0" encoding="UTF-8"?>
<Exercise name="Faulty Software System Exercise">
  <Title>Faulty Software System</Title>
  <Description> <p> After cafrefully investigating a...</p></Description>
  
  <Task type="PickVariables" feedbackLevels="0">
    <Title>Step 1: Choosing suitable propositional variables</Title>
    ...
    <Output>VARIABLES</Output>
  </Task>      
  
  <!-- State formulas for the statements -->
  <Task type="CreateFormulas" feedbackLevels="0,1,2" 
        assimilationGenerator="syntaxServer">
    <Input>VARIABLES</Input>
    <Title>Step 2: Modelling the statements</Title>
    <Description>
      Devise a formula for each observation! Use the propositional... 
    </Description>

    <Formula>
      <Description>
        If the database is faulty then so is the back end.
      </Description>
      <Solution>$D \rightarrow B$</Solution>
    </Formula>
    <Formula>
      <Description>
        The back end is only faulty if both the database and the user... 
      </Description>
      <Solution>$B \rightarrow (D \wedge U)$</Solution>
    </Formula>
    <Formula>
      <Description>Not all three components are faulty.</Description>
      <Solution>$\neg (B \wedge D \wedge U)$</Solution>
    </Formula>
                    
    <!-- Feedback generator -->
    <FeedbackGenerator> 
      <Feedback type="VariableNames">
        <Variable name="U">the user interface</Variable>
        <Variable name="B">the back end</Variable>
        <Variable name="D">the database</Variable>
      </Feedback>
    </FeedbackGenerator>
    
    <Output>FORMULAE</Output>
  </Task>
    
  <!-- State formula for the conclusion -->
  <Task type="CreateFormulas"...>
    <Input>VARIABLES</Input>
    ...
    <Output>CONCLUSIONFORMULA</Output>
  </task>

  <!-- Inferring the conclusion -->
  <Task type="CompleteFormula">
    <Input>FORMULAE</Input>
    <Input>CONCLUSIONFORMULA</Input>
    <Title>Step 4: How to infer the conclusion</Title>
    <Description>
        State an equivalence from which Julia's conclusion can be inferred.
    </Description>
    <Output>COMPLETEFORMULA</Output>
  </Task>
    
  <Task type="transformToCnf">...</task>
    <Input>COMPLETEFORMULA</Input>
    <Title>Step 5: Transformation into conjunctive normal form</Title>
    <Description>Transform the created formula to ....</Description>
    <Output>CNF_FORMULA</Output>
  </Task>
    
  <Task type="Resolution">
    <Input>CNF_FORMULA</Input>
    <Title>Step 6: Propositional resolution</Title>
    <Description>Prove that Julia's conclusion is...</Description>
  </Task>
</Exercise>
\end{lstlisting}
\end{mdframed}
\begin{center}
\begin{minipage}{.95\linewidth}
  \vspace{-5mm}
\caption{A sample XML specification of the task from Exercise \ref{example:exercise_running_example}. Outputs of some tasks are inputs to other tasks.}\label{figure:XML_specification}
\end{minipage}
 
\end{center}

\end{minipage}
\begin{minipage}{.47\linewidth}
  \begin{mdframed}
    [
      innerleftmargin=0,      innerrightmargin=0,    ]
  \flushleft \hspace{1mm} a) Translating a statement into a propositional formula 
  \begin{center}
    \vspace{-2.3mm}
    \scalebox{0.95}{\includegraphics[width=\textwidth,valign=T]{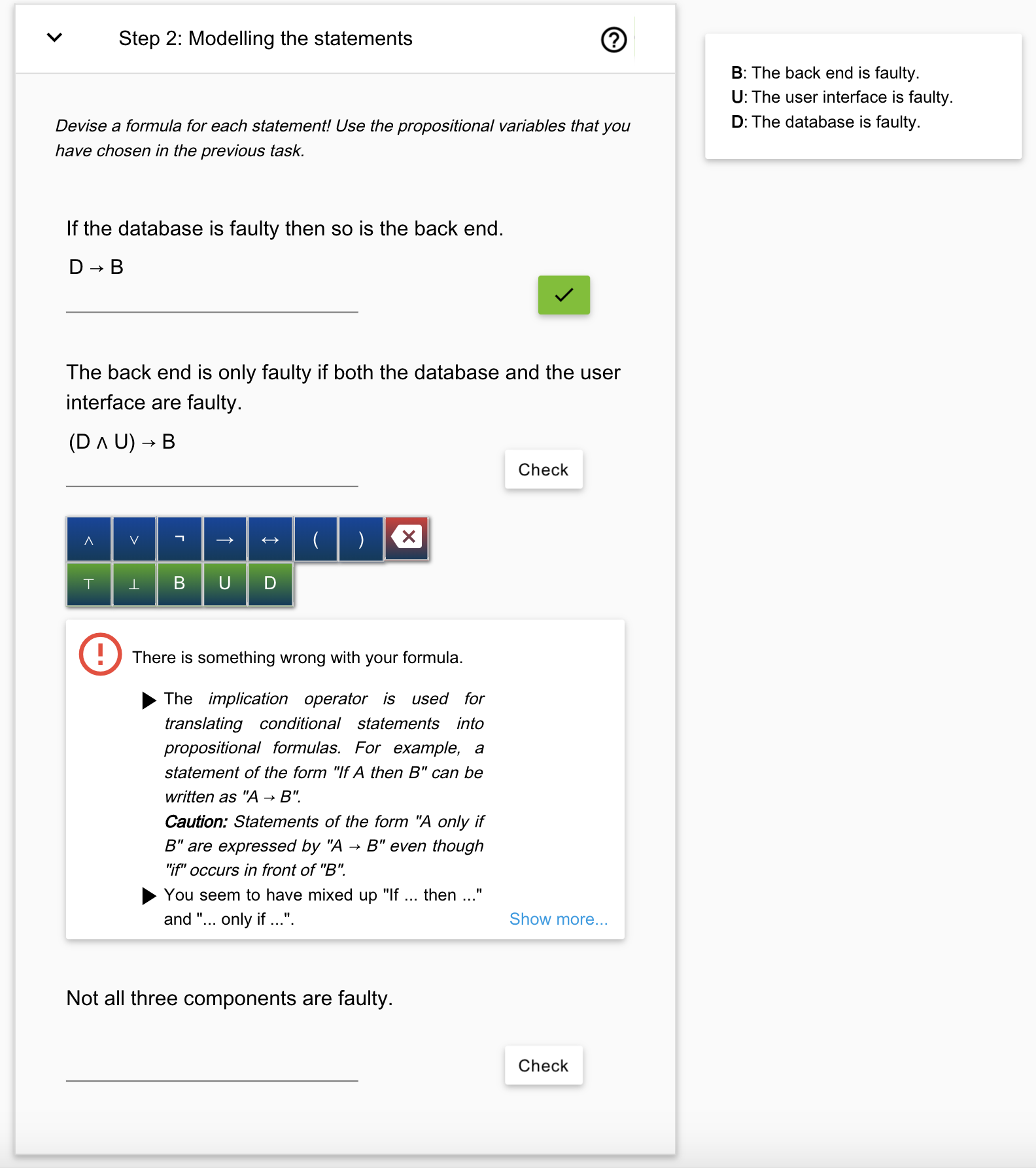}}
  \end{center}

  \flushleft \hspace{1mm} b) Transforming a propositional formula into CNF
  \begin{center}
    \vspace{-2.2mm}
    \scalebox{0.95}{\includegraphics[width=\textwidth,valign=T]{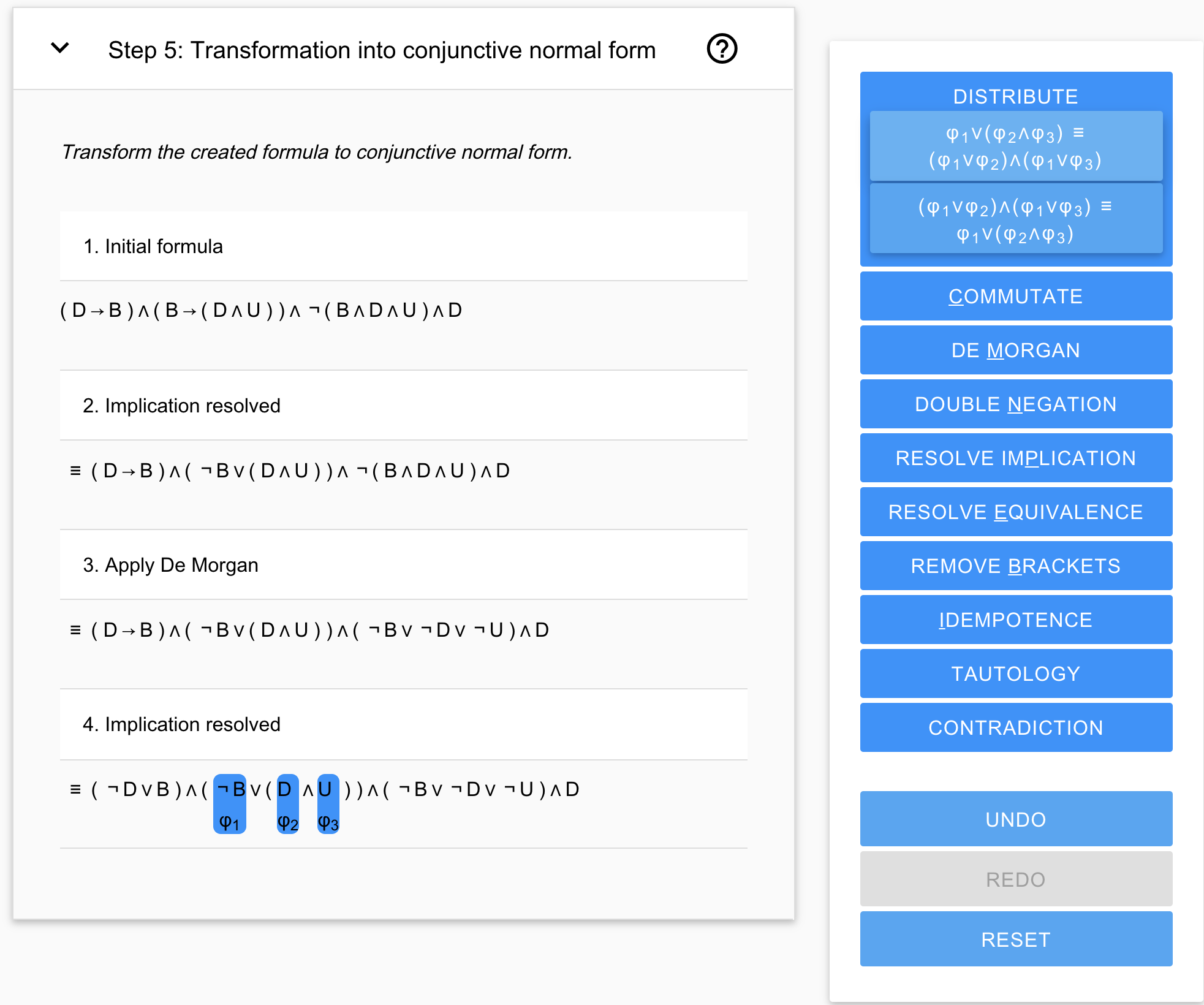}}
  \end{center}
  
  \flushleft \hspace{1mm} c) Resolving a set of clauses
  \begin{center}
    \vspace{-2.3mm}
    \scalebox{0.95}{\includegraphics[width=\textwidth,valign=T]{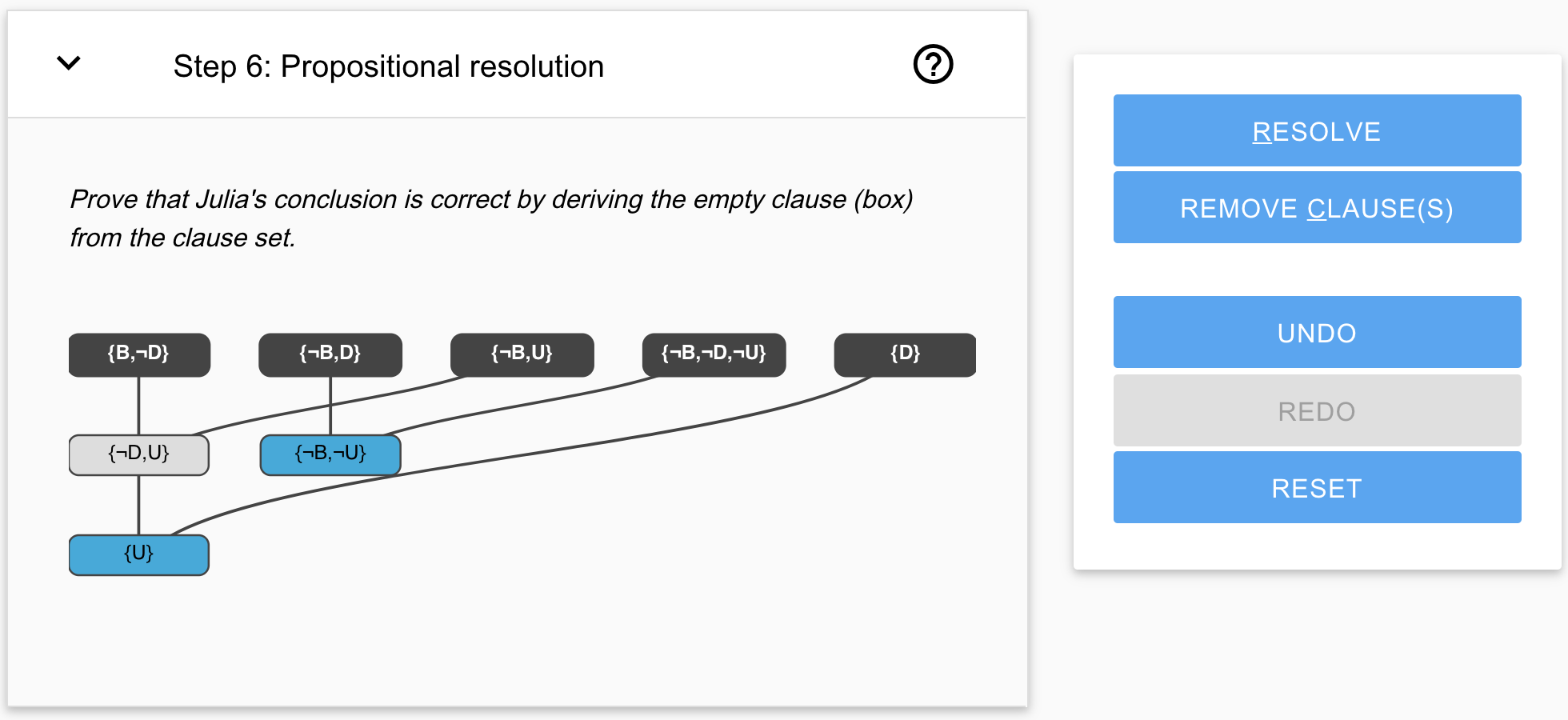}}
  \end{center}
  \end{mdframed}
\begin{center}
\begin{minipage}{.95\linewidth}
  \vspace{-5mm}
  \caption{Some of the tasks specified in the XML file from Figure \ref{figure:XML_specification} as presented by the web interface.}\label{figure:exercise_running_example}
\end{minipage}
\end{center}
\end{minipage}
\end{figure*}

The \Illtis system allows teachers to specify exercises in XML, which are then presented to students in the web (see Figures~\ref{figure:XML_specification} and \ref{figure:exercise_running_example}). Each exercise is built from one or more small tasks. As an example, the exercise described in the introduction can be built from tasks for (1) choosing suitable propositional variables, (2) formulating natural language statements as propositional formulas, (3) stating the inference goal as a satisfiability question for a formula, (4) transforming the formula into conjunctive normal form, and finally (5) applying resolution to decide satisfiability.  

So far these tasks have been implemented for propositional logic.  In addition to purely logical tasks, several helper tasks for questionnaires, for collecting feedback and data, and for starting new exercises from within an exercise have been implemented. While such tasks are helpful in designing exercises and tutorials, we focus on the more interesting logic related tasks in the following. Table \ref{table:list_of_tasks} provides an overview of the currently available tasks.

\definecolor{verylightgray}{rgb}{0.83, 0.83, 0.83}
\begin{table*}[ht]
  \scalebox{0.82}{
                                                                   \begin{tabularx}{1.2\textwidth}{|p{3.2cm}|X|p{3.95cm}|p{4.0cm}|}    \hline
    \rowcolor{lightgray}
	\textbf{Task} &
	{\centering \textbf{Description}} &
	\textbf{Input} &
	\textbf{Output} \\ \hline
	    \rowcolor{verylightgray} \multicolumn{4}{|l|}{\textbf{Logical tasks}} \\ \hline
	    PickVariable &
	Choose suitable propositional variables from a list. &
	--- &
	variables $A_1, \ldots, A_m$ \\ \hline
	    CreateFormula &
	Translate statements into formulas. &
	variables  $A_1, \ldots, A_m$ &
	formulas $\varphi_1, \ldots, \varphi_k$ \\ \hline
	    InferenceFormula &
	Combine formulas $\varphi_1, \ldots, \varphi_k$ and a formula $\varphi$ into a formula $\psi$ that is unsatisfiable if and only if $\varphi_1, \ldots ,\varphi_k$ imply $\varphi$. &
	formulas $\varphi_1, \ldots, \varphi_k$ and $\varphi$ &
	a formula $\psi$ \\ \hline
	    ManualTransformation &
	Textfield-based transformation of a formula $\varphi$ into conjunctive, disjunctive or negation normal form, or into another formula. &
	a formula $\varphi$ &
	the transformed formula $\psi$ \\ \hline
	    GuiTransformation &
	Same as previous, but graphical user interface. &
	a formula $\varphi$ &
	the transformed formula $\psi$  \\ \hline
	    Resolution &
	Resolve the empty clause from the clauses of the CNF of ~$\varphi$. &
	a formula $\varphi$ &
	---  \\ \hline
	   \rowcolor{verylightgray} \multicolumn{4}{|l|}{\textbf{Administrational tasks}} \\ \hline
   \multicolumn{4}{|l|}{
	   Questionaire task (ask a list of multiple choice questions),
	   tasks to display messages, and a task to collect data and feedback from students.
	}
   \\ \hline
  \end{tabularx}
  }
  
  \caption{List of currently implemented tasks for propositional logic.}
  \label{table:list_of_tasks}
\end{table*}

A teacher can create a new exercise by specifying a sequence of tasks in an XML format. The specification of a task includes the input and output of a task, and thereby allows to connect interdepending tasks. For example, consider a modelling task, followed by a transformation task. In the modelling task, the student is asked to provide a formula~$\psi$ for a natural language statement. The statement and a solution formula~$\varphi$ forms the input of the modelling task and, if equivalent to the solution formula, $\psi$ forms its output. The following transformation task in turn, receives formula~$\psi$.  See Figure~\ref{figure:XML_specification} for an example XML specification that describes the exercise from Example \ref{example:exercise_running_example}; some of the resulting tasks in the web interface are illustrated in Figure~\ref{figure:exercise_running_example}a--\ref{figure:exercise_running_example}c.

Most tasks come with several methods for providing feedback. These methods can be combined by the teacher in a
flexible way. For example, when a student translates a statement into a formula, the system can provide feedback on
(1) whether the formula is correct, and (2) whether certain propositional variables should (not) be
used,  but it can also analyse the formula more deeply and point out, among others, (3a) the
explanation of a logical operator that has been used in a wrong way by the student, and (3b) wrong
parts of the formula. From these methods, a teacher can choose those that are adequate for the current progress of her students, e.g., extensive feedback at the beginning of a course and less extensive feedback in exercises aimed at the preparation for examinations.

\section{The \Illtis System: The Developer's Perspective}\label{section:architecture}

One of the goals of the \Illtis framework is to allow for a modular inclusion of new types of tasks. A typical developer designing a new task shall be able to focus on the functionality of the task at hand. To this end, the mechanisms to read exercises from an XML file, to create exercise objects from such a description, and the execution of the tasks stored in such an object is implemented in a generic, task-independent way.

We focus on the internals of tasks in the following. Afterwards we summarize some further aspects that are interesting from a developer's perspective.

\paragraph{Internals of Tasks}

All tasks follow the model-view-controller pattern. In order to implement a new task, a developer has
to specify a task model that stores the current state of the task, and a task view that represents the task in the graphical user
interface; a task controller is generated automatically. In addition the developer has to specify task specific actions and feedbacks.

Upon user input, the task view generates an action that is sent to the controller. The task
controller executes the action, if it is applicable, and then returns feedback to the task view
which is then displayed.  For example, when a user does a resolution step in the resolution task, a
resolution step action is generated. The execution of this action, triggered by the controller, includes the verification that two
clauses are selected, that they can be resolved, and -- if so -- the modification of the resolution graph stored
in the resolution task model. The task model then triggers an update of the view.

The feedback is created by feedback generators. A specific task can have several feedback generators, that are arranged in a cascading fashion. For example, the task for translating statements into formulas can have two feedback generators, one for checking whether only available variables are used and one for determining where a given formula is wrong (see Section \ref{section:feedback} for more details). Developers provide implementations of the feedback generators, which can then arranged dynamically by teachers in exercise XMLs.

\paragraph{Other Features}

The framework supports anonymous data collection. All tasks come with specific loggers that have been used to collect the data for the evaluation presented in Section~\ref{section:evaluation}.
Internationalization is supported and general texts are currently provided in German and English.  
\section{Case study: Feedback for Stating Propositional Formulas}\label{section:feedback}
Feedback is essential for the learning process. Even a simple yes/no-answer gives students a feeling of their skills and allows them to seek help if necessary. All tasks implemented in \Illtis provide this rudimentary feedback. Yet, the feedback mechanism allows for more nuanced feedback.

As an example of how to use the feedback mechanism, we describe its instantiation for the task where a student shall translate a statement to a propositional formula. In this task, the propositional variables to be used as well as their intended meaning is provided by the task statement. The feedback mechanism receives a correct propositional formula $\varphi$ for the natural language statement, henceforth called \emph{solution formula}, and the formula $\psi$ provided by a student, called the \emph{student formula}.

Rudimentary feedback can be obtained by checking whether $\varphi$ and $\psi$ are logically equivalent, and, if they are not, providing an assignment that distinguishes the student formula from the solution formula. Also checks such as  verifying that only available and necessary variables are used can be performed easily.

A deeper analysis of typical errors is necessary for more meaningful feedback. In a preliminary study, we collected typical errors made by students in a final written examination. The collected errors include, among others, the use of wrong logical operators, interchanging antecedent and consequent of implications (in particular in \enquote{only if} statements as the one in statement (2) of Example~\ref{example:exercise_running_example}) and not modelling parts of the natural language statement. Furthermore, all typical errors can be found in diverse combinations.

We used these error types to implement a more advanced feedback mechanism. The idea of the mechanism is to try to transform the student formula $\psi$ into a formula $\varphi'$ that is equivalent to the solution formula $\varphi$ by reverting student mistakes. To this end, we extracted general declarative reversion rules from the list of typical errors. Each such rule searches for a pattern in $\psi$ that might result from a mistake, and transforms $\psi$ locally around this pattern.

\begin{example}
   As an example, the error where a student interchanged antecedent and consequent is specified by the reversion rule $\rho: \$X\implies \$Y \rightsquigarrow \$Y \implies\$X$. Here, the pattern $\$X \implies \$Y$ searches for a subformula which is an implication and assigns its antecedent and consequent to $\$X$ and~$\$Y$, respectively. The right hand side of the rule $\rho$ specifies how the error can be reverted, in this case by swapping $\$X$ and $\$Y$.

   When a student erroneously wrote $\psi = (D \wedge U) \implies \neg B$ instead of $\varphi =  \neg B \implies (D \wedge U)$, the rule $\rho$ is applied subsequently to all subformulas that match the left hand side of $\rho$.  As the interchanging of  antecedent and consequent is the only mistake, one of those applications yields a formula equivalent to the solution formula.

   From the type of the reversion rule as well as from its match in~$\psi$, more meaningful feedback can be constructed (see Figure \ref{figure:exercise_running_example}a).\qed
\end{example}

In general, a student can make several mistakes $m_1, \ldots, m_k$ in one formula. In this case applying the corresponding reversion rules $\rho_1, \ldots, \rho_k$ yields a formula equivalent to the solution formula.

Of course, the computational resources necessary to try to find $k$ such rule applications to the
(possibly various) matching locations grow exponentially in $k$. Fortunately, typical formulas are
very small, and therefore a rule in most cases matches at very few locations. Further, if a student made more than two mistakes in a small formula, the information that the formula is wrong is usually more useful than more precise feedback. We determined experimentally that restricting the length of reversion sequences to two yields sufficient performance, while still providing good feedback.

For the feedback in Figure \ref{figure:exercise_running_example}a the feedback generator was para\-metrised such that feedback for wrong formulas is generated and displayed in the following order:
\begin{enumerate}
  \item ``The formula is wrong.''
  \item If the formula is syntactically wrong: ``Please enter a propositional formula''.
  \item If the formula is semantically wrong:
    \begin{enumerate}
      \item If a sequence of reversion rules that yields a correct formula is found:
        \begin{enumerate}
          \item A general description of a probable misconception is displayed. For example, if the antecedent and consequent are interchanged, then an explanation of how ``If...then...'' and ``...only if...'' statements are modelled by implications is displayed.
          \item A precise description of the error is displayed. For example: ``You seem to have interchanged `If...then...' and `...only if...'\,''.
          \item The wrong part of the formula is highlighted.
        \end{enumerate}
      \item If no sequence of reversion rules could be found: An example assignment that distinguishes the student formula from the solution formula is displayed.
    \end{enumerate}
\end{enumerate}
The feedback thus provided usually gives a good idea of how to improve a formula.

Naturally, the feedback is not the best possible. A more advanced feedback mechanism could take advantage of other available information. By annotating subformulas of the  solution formula by information about the natural language formulations, the feedback can be made more precise. In the long run, we plan to model the state of learning for each student, and thereby allow for individualized feedback generation. The current implementation is the foundation for both of these extensions.
 
\section{First Classroom Experiences}\label{section:evaluation}

In the winter term 2017/2018 the \Illtis system has been used in the introductory course \enquote{Logic for Computer Scientists} at \anonymize{the Technical University Dortmund}{\anon}, aimed at second year students.

The course has fourteen weeks which are almost evenly distributed between propositional, modal and first-order logic. In addition to a two hour lecture per week, the course includes bi-weekly two hour exercise groups in which students solve problems and a weekly two-hour tutorial session for students with a need for assistance. Solutions to exercises have to be handed in by students every two weeks for being admitted for final written examinations.

The \Illtis system was used in a voluntary web-tutorial during has the part on propositional logic. We also evaluated the feedback for the modelling task described in Section \ref{section:feedback} systematically. 
\paragraph{Accompanying Web-Tutorial} An interactive tutorial covering (i) the basic connectives and
their use for modelling real world scenarios, (ii) propositional equivalences and normal forms, as
well as (iii) satisfiability in propositional logic has been published in parallel to the lectures.
Each part of the tutorial includes an introduction to the topic, and small hands-on exercises that
are checked immediately by the system. Also exercises of the full modelling process have been
included into the tutorial. At the time of writing this article, the tutorial has been accessed >700

\begin{SCtable*}
  \scalebox{0.82}{
  \begin{tabularx}{13.45cm}{|p{0.2cm} p{2.5cm}|p{0.47cm}|p{0.47cm}|p{0.47cm}|p{0.47cm}|p{0.47cm}|p{0.47cm}|p{0.47cm}|p{0.47cm}|p{0.47cm}|p{0.47cm}|p{0.47cm}|p{0.47cm}|}    \hline
    \rowcolor{lightgray}\multicolumn{2}{|l|}{\textbf{Group}} & \multicolumn{3}{c|}{\textbf{CG}} & \multicolumn{3}{c|}{\textbf{EG1}} & \multicolumn{3}{c|}{\textbf{EG2}} & \multicolumn{3}{c|}{\textbf{EG3}} \\ \hline
     \rowcolor{lightgray}\multicolumn{2}{|l|}{\textbf{Exercise}} & \multicolumn{1}{c|}{\centering \textbf{1}} & {\centering \textbf{2}} & {\centering \textbf{3}}  & {\centering \textbf{1}} & {\centering \textbf{2}} & {\centering \textbf{3}} & {\centering \textbf{1}} & {\centering \textbf{2}} & {\centering \textbf{3}} & {\centering \textbf{1}} & {\centering \textbf{2}} & {\centering \textbf{3}} \\ \hline
     \rowcolor{verylightgray} \multicolumn{14}{|l|}{\textbf{Only-if statement}} \\
     & error rate & 0.44  & 0.49 & 0.32 & 0.53 & 0.47 & 0.23 & 0.43 & 0.49 & 0.44 & 0.45 & 0.49 & 0.37\\
     & most frequent error & 0.19  & 0.30 & 0.05 & 0.33 & 0.42 & 0.02 & 0.28 & 0.41 & 0.17 & 0.33 & 0.41 & 0.08\\
     \rowcolor{verylightgray} \multicolumn{14}{|l|}{\textbf{Either-or statement}} \\
     & error rate & 0.47  & 0.51 & 0.29 & 0.58 & 0.42 & 0.07 & 0.47 & 0.38 & 0.27 & 0.50 & 0.44 & 0.17\\
     & most frequent error & 0.27  & 0.29 & 0.18 & 0.26 & 0.35 & 0.05 & 0.15 & 0.20 & 0.12 & 0.17 & 0.25 & 0.06\\  \hline
  \end{tabularx}
  }
  \caption{\small Results of the evaluation for the translation of natural language statements into propositional formulas. See Section \ref{section:evaluation} for the experimental set-up. The error rate and the rate of the most frequent error are the percentage of students of the group that made a mistake when translating the statement and the percentage of students that made the most common error, respectively.}
  \label{table:evaluation_results}
\end{SCtable*}

\paragraph{Evaluation of the Modelling Task} The feedback provided for the modelling task, as
described in the previous section, was evaluated experimentally. In their first exercise
session on propositional modelling, the students were partitioned into four groups, each of which
had to solve the same three web-based exercises. Each exercise consisted of four statements to be
modelled by propositional formulas. All three exercises contained statements of the same type, e.g.,
all of them contained an ``only...if'' and an ``either...or'' statement.

In the first and last exercise the knowledge of all students was tested, i.e.~they modelled the statements without receiving help or feedback. In the second exercise, help and feedback provided by the system differed with the group of students. The control group (CG, $n_{\text{CG}} = 57$) received no feedback; the first experimental group (EG1, $n_{\text{EG1}} = 43$) received feedback provided by the system (as described in Section~\ref{section:feedback}); the second experimental group (EG2, $n_{\text{EG2}} = 98$) was presented with a short repetition of how to model statements using propositional logic; and the third experimental group (EG3, $n_{\text{EG3}} = 51$) received the same repetition as well as feedback. 
To study the impact of help and feedback, the system logged all errors made by the students. From this data we extracted the error rate of each group for each statement type and each exercise. 
Before outlining the results we discuss some shortcomings of the set-up, and remark that the results should be taken with caution. First, the natural language statements of the three exercises are necessarily different. Therefore error rates even for statements of the same type may differ significantly over the three exercises. Further, for simplicity of the set-up, all students from a given exercise group of the logic course were assigned to the same group in the experiment. We cannot rule out that there are dependencies between students of the same exercise groups (e.g., students with mathematics as minor could be clustered in one of the groups etc.). In addition, the level of detail of feedback is not taken into account by the experiment; it would be interesting to see how detailed feedback compares to only providing students with whether their answer was correct or wrong. We plan to repeat the study in the future and to account for these problems.

For the discussion of the results, we focus on the data for the ``only...if'' and ``either...or'' statements. For each of these statements, Table \ref{table:evaluation_results} shows the error rate of each group with respect to each of the exercises. In addition, the error rate of the most common error is depicted. For both the ``only...if'' and ``either...or'' statement the most common errors (swapping antecedent and consequent/using ``or'' instead of ``either...or'') were the same over all groups and exercises. For these two errors, the system provides feedback as explained in detail in Section \ref{section:feedback}.

After receiving feedback in the second exercise, EG1 and EG3 perform much better in the third exercise. For example, the error rate for the ``only...if'' and ``either...or'' statements for EG1 drops from 0.47 and 0.42 to 0.23 and 0.07, respectively. The error rate of the most frequent error even drops from 0.42 and 0.35 to 0.02 and 0.05. As discussed above, this could be due to the different phrasings of the statements. However, the drop of the error rates of these two groups is larger than of the other two groups CG and EG2.

The results suggest that students learn from the feedback provided by the system. In a questionnaire after the evaluation, 74.6 percent of the students evaluated the system as good or very good.

\section{Conclusion and Vision}\label{section:conclusion}
The current state of the \Illtis project is a first step towards supporting students to learn logics in a modern, interactive way -- at home, in trains or at the beach. While the modelling process for propositional logic is implemented, a lot of challenges remain.

\paragraph{Extension of the System}
From a students perspective it is important to include more advanced logical formalisms as soon as possible. Currently we work on the inclusion of modal logic and predicate logic. As a foundation, an abstract term framework has been implemented that allows for including new logical mechanisms in a simple and general way.  

For modal logics and first-order logic, the ability to illustrate models is essential. For example, it is easier for students to see why a modal formula is wrong, when they see for which structure it deviates from the solution formula. Currently we work on including a framework for illustrating and manipulating graph-like structures. 

In the long run we also plan to include other topics from computer science. Many parts of the \illtis framework are likely to be reusable. For example, regular expressions can be expressed by the abstract term framework, and the declarative feedback mechanism developed for propositional logic is likely to be easily adaptable. Also typical tasks for studying and manipulating finite state automata, as well as feedback mechanisms for those tasks,  can likely be obtained by adapting tasks performed on Kripke structures. 

The inclusion and development of techniques to increase the motivation of students is planned as well.

\paragraph{Feedback Generation}

Feedback generation for more expressive logics than propositional logics is, of course, not easy. For modal logic the declarative mechanism designed for propositional logic can likely be adapted. In addition, errors can be illustrated by counter examples as explained above.

Verifying the correctness of a first-order formula, proposed as solution by a student, cannot be performed algorithmically in general (since it is undecidable). We plan to investigate workarounds. One such workaround is to treat first-order formulas as selection mechanism (i.e. as queries). Imagine a graph in which several nodes are marked, and a student is asked to  write a formula with one free variable that selects the marked nodes. Then an error in a student formula can be easily illustrated by highlighting the set of nodes selected by the formula (which differs from the marked nodes). Another workaround is to identify a decidable fragment of first-order logic that includes typical formulas provided by students (i.e~short formulas with few quantifications and quantifier alternations).
\paragraph{Didactical Research}

The \Illtis project offers the opportunity to study how students learn logic. For example,
anonymised data can be collected about the types of errors made when constructing formulas, and the
amount of time spent on each formula. The analysis of such data will likely lead to a better understanding of typical misconceptions of students.

\paragraph{Acknowledgements}
We thank \anonymize{Nils Vortmeier}{\anon} for the collection of an initial data set of student mistakes. We are grateful to \anonymize{Johannes Fischer}{\anon},  \anonymize{Marko Schmellenkamp}{\anon}, and \anonymize{Thomas Schwentick}{\anon} for valuable feedback on a draft of this article.

\bibliographystyle{apalike}
\bibliography{bibliography}

\end{document}